\documentclass[prd,amsmath,amssymb,superscriptaddress,floatfix,nofootinbib,10pt]{revtex4}
\usepackage{times}
\usepackage{amssymb,amsbsy,amsmath,amsfonts}
\usepackage{graphicx}
\usepackage{float}
\usepackage{color}
\usepackage{morefloats}
\usepackage{rotating}
\usepackage{srcltx}
\usepackage{slashed}
\usepackage{multirow}
\usepackage{verbatim}
\usepackage{hyperref}
\usepackage{tabularx}
\usepackage{bm}
\allowdisplaybreaks[4]

\usepackage{bbding}
\usepackage{threeparttable}

\DeclareUnicodeCharacter{2212}{\ensuremath{-}}

\newcommand{\PreserveBackslash}[1]{\let\temp=\\#1\let\\=\temp}
\newcolumntype{C}[1]{>{\PreserveBackslash\centering}p{#1}}
\newcolumntype{R}[1]{>{\PreserveBackslash\raggedleft}p{#1}}
\newcolumntype{L}[1]{>{\PreserveBackslash\raggedright}p{#1}}

\begin{document}

\title{ Role  of $a_0( 1710)$ in the $D_s^+\to\rho^+\phi,~~\rho^+\omega$ decays }

\author{Jing Song}
\affiliation{School of Physics, Beihang University, Beijing, 102206, China}
\affiliation{Departamento de Física Teórica and IFIC, Centro Mixto Universidad de Valencia-CSIC Institutos de Investigación de Paterna, 46071 Valencia, Spain}

\author{Zi-Ying Yang}
\affiliation{School of Physics, Beihang University, Beijing, 102206, China}
\affiliation{Departamento de Física Teórica and IFIC, Centro Mixto Universidad de Valencia-CSIC Institutos de Investigación de Paterna, 46071 Valencia, Spain}

\author{ Eulogio Oset}
\email[]{oset@ific.uv.es}
\affiliation{Departamento de Física Teórica and IFIC, Centro Mixto Universidad de Valencia-CSIC Institutos de Investigación de Paterna, 46071 Valencia, Spain}
\affiliation{Department of Physics, Guangxi Normal University, Guilin 541004, China}

\begin{abstract}
We look into the $D_s^+ \to \rho^+ \phi$ and $D_s^+ \to \rho^+ \omega$ weak decays recently measured by the BESIII collaboration, which proceed very differently in a first step of the weak decay. While the first reaction proceeds directly via external emission, the second one does not go via external nor internal emission, what prompted the experimental team to claim that it proceeds via $W-$annihilation. We show in this work that the unexpectedly large rate of the  
$D_s^+ \to \rho^+ \omega$, should it be due to $W-$annihilation, has a different explanation since is it naturally obtained once final state interaction of the $\rho^+ \phi, ~\rho^+ \omega$ and $K^{*+} \bar{K}^{*0}$ channels is taken into consideration. The interaction of these channels produces the $a_0(1710)$ resonance, predicted long ago as a molecular state of these coupled channels, and only recently observed, and it is the presence of this resonance what makes the effect of the final state interaction very important in the weak decays and provides a natural explanation of the experimental decay rates. 
\end{abstract}

\maketitle
\section{Introduction}

The study of the vector vector ($VV$) interaction was initiated in the works of \cite{Molina:2008jw}, with the interaction of $\rho \rho$, and \cite{Geng:2008gx} with the extension to the $SU(3)$ space. Many resonances are obtained dynamically generated from this interaction which is obtained in the framework of the  local hidden gauge approach \cite{Bando:1984ej,Bando:1987br,Meissner:1987ge,Nagahiro:2008cv}, exchanging vector mesons. The $f_0(1370)$ and  $f_2(1270)$ resonances are obtained from the $\rho \rho$ interaction, while extending the theory to  include the strange sector, many other resonances are obtained, as the  $f_0(1710), f'_2(1525)$, and $K^*_2(1430)$. The decay channels into two pseudoscalars ($PP$) are considered in \cite{Molina:2008jw} and \cite{Geng:2008gx} through box diagrams. In \cite{Wang:2019niy} the $PP$ channels are considered as coupled channels together with the $VV$ ones for the $\rho  \rho$ case, with results similar to those of \cite{Molina:2008jw}. 

  In \cite{Geng:2008gx} predictions were made for a new $a_0$, scalar $I=1$ resonance, with mass around 1777 MeV, that was recently vindicated by several experimental works,  Ref.~\cite{BaBar:2021fkz}  in Babar, Ref.~\cite{BESIII:2021anf}  in BESIII and Ref.~\cite{LHCb:2023evz} in LHCb. The resonance is already cataloged in the PDG as the $a_0(1710)$~\cite{ParticleDataGroup:2024cfk}. The masses and widths reported by the different experiments are shown in Table~\ref{table1_1}.  
 \begin{table}[H]
\centering
\caption{{ Masses and widths for the $a_0 (1710)$ (in units of MeV).}}
\label{table1_1}
\setlength{\tabcolsep}{48pt}
\begin{tabular}{ccc}
\hline \hline
&  { Mass [MeV] } & $\Gamma[\mathrm{MeV}]$ \\
\hline 
 { PDG~\cite{ParticleDataGroup:2024cfk} } & $1713 \pm 19$ & $107 \pm 15$ \\
 { LHCb~\cite{LHCb:2023evz} } & $1736 \pm 10 \pm 12$ & $134 \pm 17 \pm 61$ \\
{ BESIII~\cite{BESIII:2022npc} } & $1817 \pm 8 \pm 20 $& $97 \pm 22 \pm 15$ \\
 { Babar~\cite{BaBar:2021fkz} } & $1704 \pm 5 \pm 2$ & $110 \pm 15 \pm 11$ \\
 { Geng et al.~\cite{Geng:2008gx} } & $1777$ & $148$ \\
 { Du et al.~\cite{Du:2018gyn} } & $1750-1790$ & no calculated width \\
\hline
\hline
\end{tabular}
\end{table}
  As one can see, there are still some discrepancies about the precise value of the mass, but there is consensus about the existence of the resonance and the value of the width, between 100 and 150 MeV. In the Table we also show the prediction of Ref. \cite{Geng:2008gx} and the results of a recent independent calculation for the mass, using dispersion relations \cite{Du:2018gyn}.  As one can see, there is agreement about the mass between these two independent calculations. Similar results, depending somewhat on the cut off chosen, are obtained in \cite{Wang:2022pin}. 

   The experimental observation of the $a_0$ state had immediate theoretical repercussion.  In the molecular picture of \cite{Geng:2008gx,Du:2018gyn} the $a_0(1710)$ couples to the $K^*\bar{K}^*$ in $I=1$ and to $\rho \phi, ~\rho \omega$.
In \cite{Dai:2021owu}  the $D_s^+ \to \pi^+K^{*+}K^{*-},~ \pi^+K^{*0}\bar{K}^{*0}$ reactions were studied and the final state interaction of $K^* \bar{K}^*$ was studied by means of the $f_0(1710)$ and $a_0(1710)$ dynamically generated resonances. The vector-vector ($VV$) channels were allowed to make a transition to the  $K^+K^−, ~K^0
 \bar{K}^0$ and with this mechanism a ratio of the decay rates of these two  decay channels was found, in agreement with experiment.  In the same work a suggestion was made for a reaction which would isolate the $a_0(1710)$ resonance, the   $D^+_s \to \pi^0K^+K_S$ reaction, making predictions for the branching ratio of the reaction. This reaction was later observed in \cite{BESIII:2022npc} with results similar to those predicted in \cite{Dai:2021owu}. 
 In Ref. \cite{Zhu:2022wzk} the authors retake the idea of \cite{Dai:2021owu} and show that the consideration of the $f_0(1710)$ and $a_0(1710)$ resonances produces mass distributions for $K^0_SK^0_S$ and $\pi^+K_0^S$  invariant mass distributions in agreement with experiment. 
 In Ref. \cite{Zhu:2022guw} the $D^+_s \to a_0^+(1710)\to \pi^0K^+K_S^0$ reaction is studied and, in addition to $a_0(1710)$, the contributions of $K^*$ and $a_0(980)$ are also taken into account. 
 It is found that the experimental measurements on the $K^+K^0_S,~\pi^0K^+$, and $\pi^0K^0_S$ invariant mass distributions can be well
reproduced, which supports the molecular, mostly $K^*\bar{K}^*$, nature of the scalar $a_0(1710)$ resonance. In \cite{Abreu:2023xvw}  the $J/\psi \to \phi K \bar{K}$ decay is studied, looking for differences in the production rates of $K^+K^−$ or $K^0\bar{K}^0$
in the region of 1700-1800 MeV, where the two resonances $f_0(1710)$ and $a_0(1710)$ appear, and predictions are made for mass distributions. In 
\cite{Wang:2023aza}  the $D^+_s \to K^0_SK^+\pi^0$ decay is again reanalyzed, and the
contributions of the resonances $a_0^+(980),~ a_0^+(1710)$ in  $S-$wave and $\bar{K}^{*0}(892), {K}^{*+}(892)$ in  $P-$wave are considered, and agreement with  experimental data is also shown.  In Ref. \cite{Achasov:2023izs} the authors  advocate for the measurement of land-shapes in $a_0(1700/1800) \to K^*\bar{K}^*, ~\rho\phi, ~\rho\omega$ decays as a method to further learn about these $VV$ dynamically generated resonances. In Ref. \cite{Wang:2023lia} the authors investigate the  $K^- p \to  a_0^+(1710)$ reaction as a means of learning more about this $a_0$ resonance, evaluating cross sections and suggesting the optimal energies where this resonance plays a relevant role. In \cite{Ding:2023eps} the authors suggest to look for the  the $a_0(1710)$  resonance in the $\eta_c \to  \bar{K}^0K^+\pi^-$ 
reaction. In Ref. \cite{Ding:2024lqk} the molecular picture is again retaken and predictions are made for the $J/\psi \to \bar{K}^0 K^+\rho^-$  reaction, stressing the role payed by the  $a_0(1710)$ resonance. In \cite{Peng:2024ive} the $a_0(1710)-f_0(1710)$ mixing effect in the $D_s^+\to K_S^0K_S^0 \pi+$ decay is studied in detail considering the experimental mass distributions. A perspective on these and other possible reactions concerning the $a_0(1710)$ resonance is provided by the paper \cite{Oset:2023hyt}.

Very recently the role of the $f_0(1710)$ and $a_0(1710)$ resonances in the $D^0 \to \rho^0 \phi$, $\omega \phi$ decays was established. These two decays proceed in a direct mode via internal emission with equal rates.
Yet, the experimental branching ratio for the $\rho^0\phi$ mode is twice as big as that for the $\omega \phi$ mode \cite{ParticleDataGroup:2024cfk,Cao:2023csx}. By considering the contribution of the  $f_0(1710)$ and $a_0(1710)$ resonances in these decays, a ratio for the branching fractions consistent with experiment was found. 

    The present work addresses the production of $\rho^+ \phi$ and $\rho^+ \omega$ in $D_s^+$ decay which has been recently measured by the BESIII collaboration \cite{BESIII:2025owp}. The reactions are interesting because the $\rho^+ \phi$ mode appears immediately via external emission, the strongest of the weak decay modes  \cite{Chau:1982da}, while the $\rho^+ \omega$ mode is not produced directly neither via external nor internal emission. The BESIII team in \cite{BESIII:2025owp} interpreted this fact suggesting that the  $\rho^+ \omega$ decay mode proceeds via $W$-annihilation. This would provide a $W-a$nnihilation decay mode with an exceptionally large decay rate since experimentally it is found in \cite{BESIII:2025owp} that the $\rho^+ \omega$ decay rate is only a factor four smaller than that of the $\rho^+ \phi$ mode. This is indeed surprising since the strength of the weak decays decreases in the order \cite{Chau:1982da}: external emission, internal emission, $W -$exchange, $W -$annihilation, horizontal $W -$loop and vertical $W -$loop, and only from external emission to internal emission one already expects an order of magnitude decrease in the decay rates. However, we shall see that one does not need to invoke $W-a$nnihilation. We can obtain the right ratio of decay rates by means of external plus internal emission, together with final state interaction. It is important to recall from the beginning that in order to have a given final state in the weak decay one does not need to produce this state in a first step. It is sufficient to produce another channel which is coupled to the final one by means of strong interaction, and then allow the transition from one channel to the other by means of the strong interaction rescattering. This is not the first time that this occurs: Indeed, in \cite{BESIII:2019jjr} the $D^+_s \to a^+_0(980)\pi^0$ and $D^+_s \to  a^0(980)\pi^+$  decays leading to  $D^+_s \to \pi^+\pi^0 \eta$ were reported and the mechanism was branded as a clean example of $W-$annihilation, with a rate which
is one order of magnitude bigger than the typical $W-$annihilation rates.  
This, however, was explained by means of a mechanism of internal emission via the $D_s^+ \to K^+ K^- \pi^+, ~D_s^+ \to K^0  \bar{K}^0 \pi^+$ followed by final state interaction of $K \bar{K} \to  \pi \eta$ \cite{Molina:2019udw}, which is driven by the $a_0(980)$ resonance, dynamically generated from the interaction  of the $K \bar{K}$ and $\pi \eta$  channels \cite{Oller:1998hw} (see related works on the subject in~\cite{Hsiao:2019ait,Ling:2021qzl,Bayar:2023azy}). Another example is found in the $D^+_s \to \pi^+\pi^+\pi^-\eta$ reaction, measured for the first time in~\cite{BESIII:2021aza}  by the BESIII collaboration. One surprising aspect of the analysis of~\cite{BESIII:2021aza}  is the claim that the $D^+_s\to  a^+_0(980)\rho^0$  decay mode proceeds via weak annihilation, with a rate about one order of magnitude bigger than ordinary weak-annihilation processes. Once again it was shown in the detailed theoretical work of \cite{Song:2022kac} that this decays mode was unnecessary, and the 6 invariant mass distributions were well described starting from external and internal emission mechanisms followed by final state interaction.

   In the present work we shall also see that starting from the external and internal emission decay mechanisms of the $D_s^+ \to \rho^+ \phi$ and $K^* \bar{K}^*$, respectively, together with final state interaction of $VV$, we shall be able to obtain the ratio of  $D_s^+ \to \rho^+ \omega$ to $D_s^+ \to \rho^+ \phi$ decay rates compatible with the experimental observation. Since the $VV$ channels are in isospin $I=1$, the interaction of these channels generate the $a_0(1710)$ and the  $D_s^+ \to \rho^+ \omega$ decay is directly tied to this resonance. In other words, it is the presence of the $a_0(1710)$ resonance that makes the  $D_s^+ \to \rho^+ \omega$ possible and with large strength, without having to invoke the $W-$annihilation mode.

\section{formalism}
The BESIII experiment of \cite{BESIII:2025owp} reports
\begin{align}
    BR(D_s^+ \to \rho^+ \phi) &= (3.98\pm0.33\pm0.21)\times10^{-2},\nonumber\\
    BR(D_s^+ \to \rho^+ \omega) &= (0.99\pm0.08\pm0.07)\times10^{-2}. \label{1_1}
\end{align}

\begin{figure}[H]
  \centering
\includegraphics[width=0.4\textwidth]{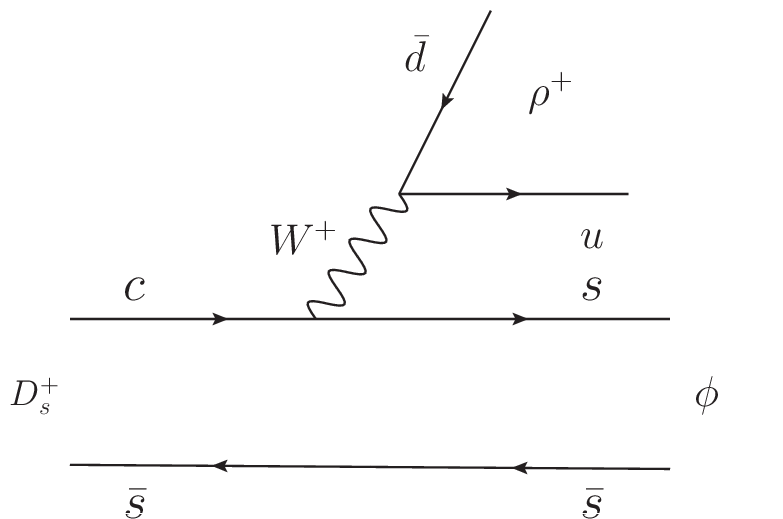}   
  \caption{Mechanism for $D_s^+ \to \rho^+ \phi$ from external emission.}\label{fig1}
\end{figure}

\begin{figure}[H]
  \centering
\includegraphics[width=0.4\textwidth]{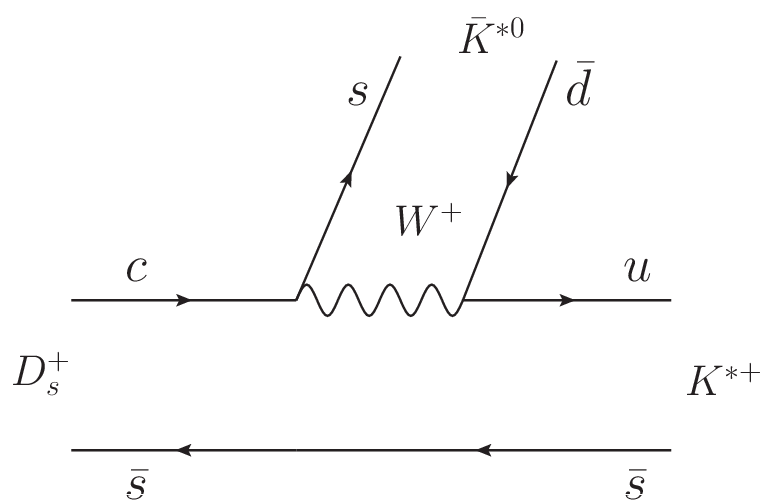}   
  \caption{Mechanism for $D_s^+ $ decay based on internal  emission.}\label{fig2}
\end{figure}

\begin{figure}[H]
  \centering
\includegraphics[width=0.4\textwidth]{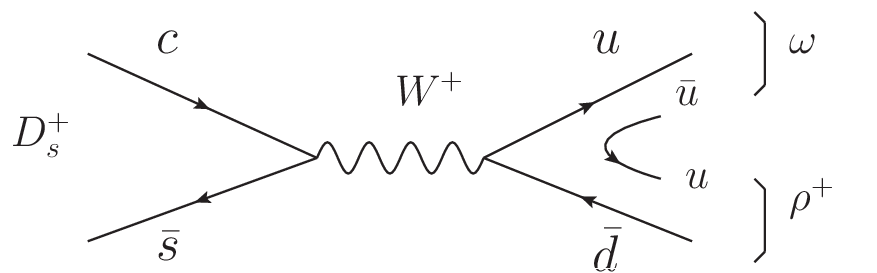}   
  \caption{Mechanism for $D_s^+ \to \rho^+ \omega$ based on $W-$annihilation as suggested in~\cite{BESIII:2025owp}.}\label{fig3}
\end{figure}

\begin{figure}[H]
    \centering
\includegraphics[width=0.28\textwidth]{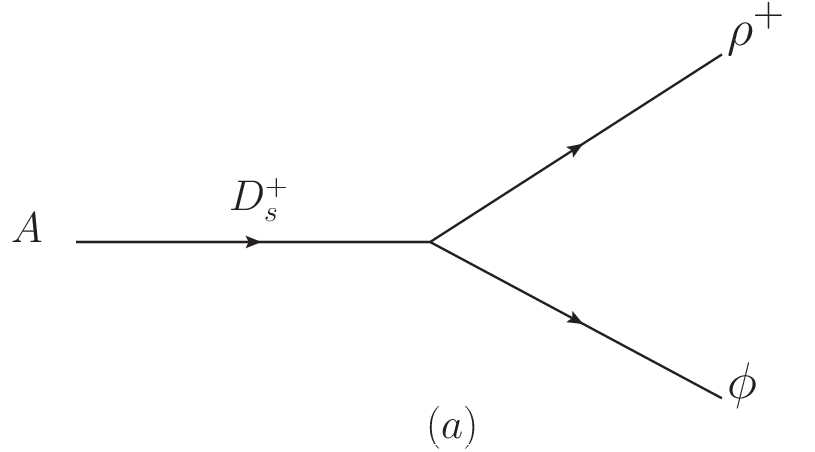}
\includegraphics[width=0.35\textwidth]{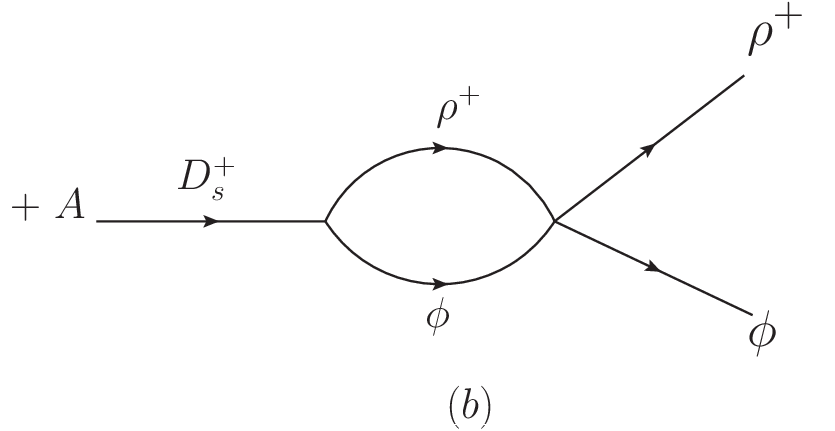}
\includegraphics[width=0.35\textwidth]{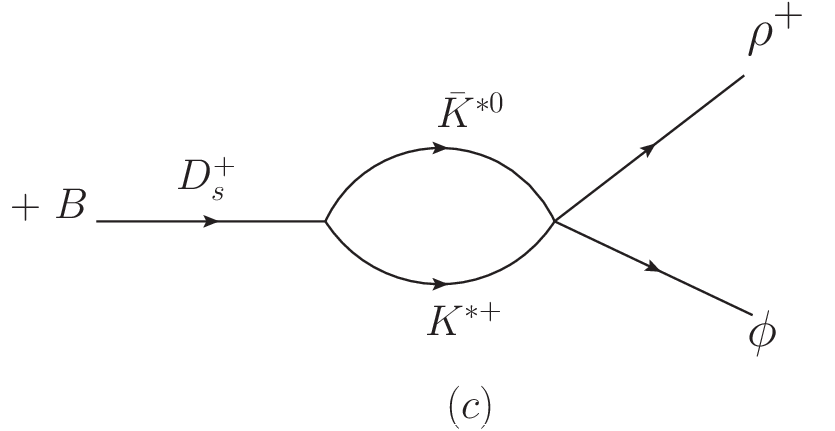}
\includegraphics[width=0.33\textwidth]{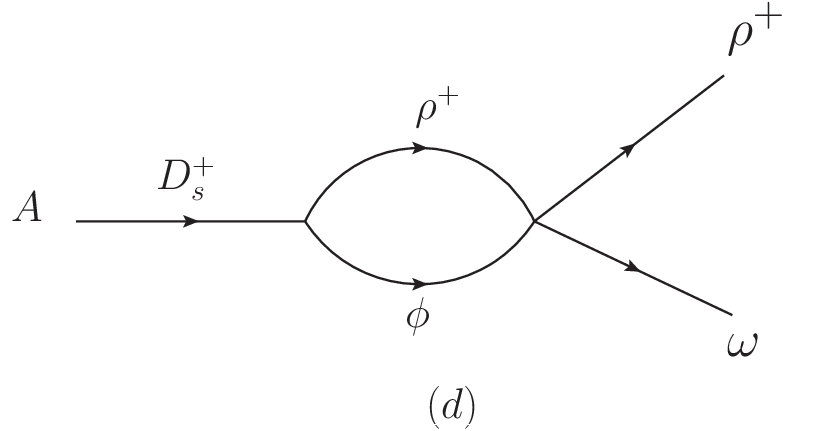}
\includegraphics[width=0.35\textwidth]{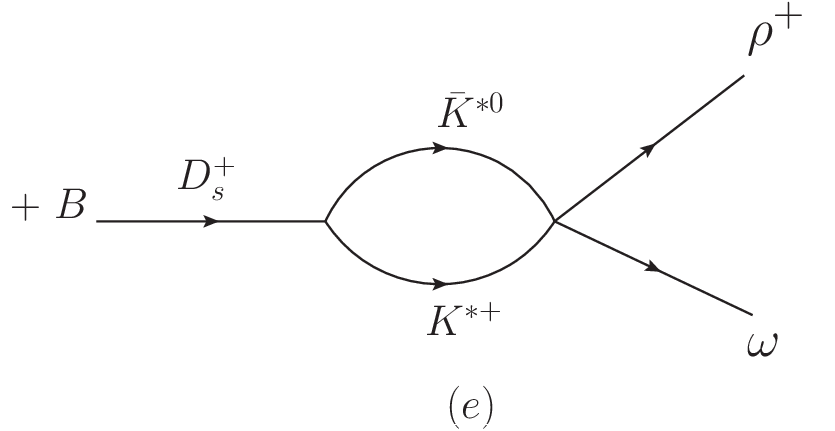}
   \caption{ Diagrams involving final state. (a) (b) (c) for $\rho^+ \phi$ production, (d) (e) for $\rho^+ \omega$ production. } \label{fig4} 
\end{figure}

Summing errors in quadrature, we obtain a ratio
\begin{align}\label{1_2}
    R_B=\frac{BR(D_s^+ \to \rho^+ \phi)}{BR(D_s^+ \to \rho^+ \omega)} = [3.5 \sim 6].
\end{align}
The $\rho^+ \phi$ production proceeds, as shown in~\cite{BESIII:2025owp} by means of external emission as shown in Fig.~\ref{fig1}.
The $s\bar{s}$ produced in Fig.~\ref{fig1}  gives rise to the $\phi$, but not to the $\omega$ which is $\frac{1}{\sqrt{2}}(u \bar{u}+d\bar{d})$. The mechanism of internal emission shown in Fig.~\ref{fig2} also does not directly produce $\rho^+\omega$, but two strange mesons. This reaction {leads} the BESIII Collaboration to claim that this production mode
proceeds via $W-$annihilation as shown in Fig.~\ref{fig3}. However, as discussed in the Introduction, and according to general values of weak decays~\cite{Chau:1982da}, one expects the ratio of this mechanism versus external emission to be suppressed by at least two orders of magnitude, in clear disagreement with the band of values of Eq.~(\ref{1_2}). In view of this, and recalling the finding of Refs.~\cite{Molina:2019udw,Song:2022kac}, where experimental claims of $W-$annihilation~\cite{BESIII:2019jjr,BESIII:2021aza} were dismissed in favor of a final state interaction mechanism, we look now at the role played by final state interaction in the  $D_s^+ \to \rho^+ \phi$ and $D_s^+ \to \rho^+ \omega$ decays.
As shown in~\cite{Geng:2008gx,Wang:2019niy,Wang:2022pin,Du:2018gyn} the $VV$ interaction in $I=1$ is driven by the channels $K^*\bar{K}^*,~\rho^+\phi$, and $\rho^+\omega$, where the $a_0(1710)$ is dynamically generated. The amplitude behaves as 
\begin{align}
&t_{ij}=\Tilde{t}_{ij}P^{(0)};\nonumber\\
    &\Tilde{t}_{ij}=\frac{g_i~g_j}{s-s_R},\qquad\qquad i,~j=K^*\bar{K}^*,~\rho^+\phi,~\rho^+\omega,
\end{align}
close to the pole, where $g_i$ are the couplings of the resonance to the coupled channels, which we reproduce in Table~\ref{table1} extracted from the work of~\cite{Geng:2008gx}. $P^{(0)}$ is the spin projection in $J=0$ of the two vectors~\cite{Molina:2008jw} 
\begin{align}
    P^{(0)}= \frac{1}{3}(\vec{\epsilon}_1\vec{\epsilon}_2)~(\vec{\epsilon}_3\vec{\epsilon}_4)
\end{align}
with $\vec{\epsilon}_i$ the vector polarization of the vector meson and the indices corresponding to $1+2\to 3+4$
\begin{table}[H]
\centering
\caption{Couplings of the $a_0(1710)$  to $K^*\bar{K}^*,~\rho^+\phi,~\rho^+\omega$ for the $I=0$ states  and charged states (in units of MeV).}
\label{table1}
\setlength{\tabcolsep}{38pt}
\begin{tabular}{cccc}
\hline \hline
\multirow{4}*{$g_i$}  &$K^*\bar{K}^*$ & $\rho\phi$ & $\rho\omega$  \\\cline{2-4}
 & $7525 - i~1529$ & $4998 - i~1872$ & $-4042 + i~1391$\\
  & $K^{*+}\bar{K}^{*0}$ & $\rho^+\phi$ & $\rho^+\omega$ \\\cline{2-4}
& $7525 - i~1529$ & $-4998 + i~1872$ & $4042 - i~1391$\\
\hline
\end{tabular}
\end{table}
Note that since we have the phase convention for isospin multiplets, $(K^{*+},K^{*0}),~(\bar{K}^{*0},-{K}^{*-}),~(-\rho^+,\rho^0,\rho^-)$, the couplings for the charge states in Table~\ref{table1} change sign for $\rho^+ \phi,~ \rho^+ \omega$ compared to those in the isospin basis.
As we can see in Fig.~\ref{fig1} we produce $\rho^+ \phi$ by external emission, to which we associated a weight $A\vec{\epsilon}_1\vec{\epsilon}_2$ in $S-$wave  and in  Fig.~\ref{fig2} we produce the vectors $\bar{K}^{*0}K^{*+}$ by internal emission, to which, we associate a weight $B\vec{\epsilon}_1\vec{\epsilon}_2,~~B=\beta A$.
We expect, based on the color counting, which favors external emission by a factor of the order of $N_c$, that $\beta{\sim}1/3$.
Considering the final state interaction, the mechanism for $\rho^+ \phi$ and $\rho^+ \omega$ are shown in Fig.~\ref{fig4}.
Analytically we will have now for $D_s^+ \to \rho^+ \phi$ and $D_s^+ \to \rho^+ \omega$, 
\begin{align}
    t_{\rho^+ \phi} = \Tilde{t}_{\rho^+ \phi}\vec{\epsilon}_{\rho^+}\vec{\epsilon}_{\phi};~~~~
    t_{\rho^+ \omega} = \Tilde{t}_{\rho^+ \omega}\vec{\epsilon}_{\rho^+}\vec{\epsilon}_{\omega},
\end{align}
with
\begin{align}\label{5_1}
     &\Tilde{t}_{\rho^+ \phi} = A~\Bigg( 1+ G_{\rho^+ \phi}(m_{D_s^+})~\Tilde{t}_{\rho^+ \phi, \rho^+ \phi}(m_{D_s^+})  +\beta~G_{K^* \bar{K}^*}(m_{D_s^+})~\Tilde{t}_{K^* \bar{K}^*,\rho^+ \phi}(m_{D_s^+})  \Bigg),  
\end{align}

\begin{align}\label{5_2}
     &\Tilde{t}_{\rho^+ \omega} = A~\Bigg(  G_{\rho^+ \phi}(m_{D_s^+})~\Tilde{t}_{\rho^+ \phi, \rho^+ \omega}(m_{D_s^+})  +\beta~G_{K^* \bar{K}^*}(m_{D_s^+})~\Tilde{t}_{K^* \bar{K}^*,\rho^+ \omega}(m_{D_s^+}) \Bigg),    
\end{align}
with the $\Tilde{t}_{ij}$ matrices given by 
\begin{align}
    &\Tilde{t}_{\rho^+ \phi,\rho^+ \phi} = \frac{g_{\rho^+ \phi}^2}{m_{D_s^+}^2-M_R^2+i M_R \Gamma_R};\nonumber\\
    &\Tilde{t}_{\rho^+ \phi,\rho^+ \omega} = \frac{g_{\rho^+ \phi}g_{\rho^+ \omega}}{m_{D_s^+}^2-M_R^2+i M_R \Gamma_R};\nonumber\\
    &\Tilde{t}_{K^* \bar{K}^*,\rho^+ \phi} = \frac{g_{K^* \bar{K}^*}g_{\rho^+ \phi}}{m_{D_s^+}^2-M_R^2+i M_R \Gamma_R};\nonumber\\
    &\Tilde{t}_{K^* \bar{K}^*,\rho^+ \omega}= \frac{g_{K^* \bar{K}^*}g_{\rho^+ \omega}}{m_{D_s^+}^2-M_R^2+i M_R \Gamma_R},
\end{align}
with the couplings from Table~\ref{table1}.
The $G$ functions from Eqs.~(\ref{5_1})~(\ref{5_2})  are the meson-meson loop function which are regularized with a cut off, given by
\begin{align}\label{cut}
   G = \int_{|\vec{q}~|<q_{\text{max}}} \frac{d^3 q}{(2 \pi)^3} \frac{  w_1(q) + w_2(q) }{2 w_1(q) w_2(q)} \frac{1}{s - (w_1(q) + w_2(q))^2 + i \epsilon},
\end{align}
with $w_i=\sqrt{\vec{q~}^2+m_i^2}$ and $m_i$ the mass of the meson, and following Ref.~\cite{Geng:2008gx,Ikeno:2024hiw} we take $q_{\text{max}}=960$~MeV.

The decay widths are given by
\begin{align}
    \Gamma_i=\frac{1}{8\pi}\frac{1}{m_{D_s^+}^2} 3
|\Tilde{t}_i|^2 q_{\text{on},i},
\end{align}
 with the factor 3 coming from the sum of the final vector polarizations, and $q_{\text{on},i}$,   given by 
\begin{align}
& q_{\text{on},i}=\frac{\lambda^{1 / 2}\left(m_{D_s^+}^2,~ m_{1,i}^2,~ m_{2,i}^2\right)}{2 m_{D_s^+}};~~~~i=\rho^+\phi,~\rho^+\omega.
\end{align}

\section{results}
In order to get the ratio of $R_B$ 
of Eq.~(\ref{1_2}), 
we only have one parameter free, which is the value of $\beta$, since the factor A cancels in the ratio, and we would expect that the value of $\beta$ is around $1/3$ or reasonably smaller than 1. Only if this is the case can we claim that we find a sensible solution to the problem. Note that if the final state interaction (FSI) had a small effect, we would not get any solution to this problem, even with unreasonable values of $\beta$. This is not the case, and the final state interaction is quite effective. To show that we first see what we get for  $\Gamma_{\rho^+\phi}$ with or without final state interaction, comparing the final $\Gamma_{\rho^+\phi}$ with the one obtained with just the tree level (term 1 in Eq.~(\ref{5_1})). By taking $\beta=1/3$ and $\beta=0$, we find
\begin{align}\label{7_2}
    \frac{\Gamma_{\rho^+\phi}(\beta=1/3)}{\Gamma_{\rho^+\phi}\text{(tree level)}} = 0.64;\nonumber\\
    \frac{\Gamma_{\rho^+\phi}(\beta=0)}{\Gamma_{\rho^+\phi}\text{(tree level)}} = 0.35.   
\end{align}
The results of Eq.~(\ref{7_2}) are obtained with the inputs of~\cite{Geng:2008gx} shown in Tables~\ref{table1_1} and \ref{table1}. Using other masses and widths from Table~\ref{table1_1} the results are qualitatively similar. What we see is that just with the effect of the FSI in external emission ($\beta=0)$, there is a reduction of the width by about a factor three. The extra effect of the FSI  due to internal emission ($\beta=1/3)$ softens the global effects of the FSI, reducing the width by about a factor $\frac{2}{3}$. This indicates that the contribution  of the internal emission is not negligible and that it has apposite sign to that of the external emission due to the signs of the couplings in  Table~\ref{table1}. The sizable strength of the internal emission mechanism is due to the nearly double strength of the $K^*\bar{K}^*$ couplings  with respect to that of $\rho^+\phi$. What we can see it that the FSI is very important and it is tied to relevant information on the $VV$ interaction in this particular case.

Once the important effect of the final state interaction is proved, we proceed to see its effect in the ratio $R_B$ of Eq.~(\ref{1_2}). We aim at obtaining $R_B\approx5$ in the middle of the experimental bracket of Eq.~(\ref{1_2}). Then we tune the value of  $\beta $ for different cases to obtain that value and we show the results in Table~\ref{table2}.
\begin{table}[H]
\centering
\caption{Different values of $\beta$ needed to obtain $R_B\approx5$.}
\label{table2}
\setlength{\tabcolsep}{42pt}
\begin{tabular}{cccc}
\hline \hline
  &Geng et al.~\cite{Geng:2008gx} & LHCb~\cite{LHCb:2023evz} & BESIII~\cite{BESIII:2022npc} \\
\hline
$\beta$ & $0.19$ & $0.1$ & $0.29$\\
\hline
\end{tabular}
\end{table}
We have taken as an example the mass and width of~\cite{Geng:2008gx}, the one of LHCb~\cite{LHCb:2023evz} and the one of the BESIII~\cite{BESIII:2022npc}, respectively,
where discrepancies in the values of mass and width still persist. In all of them we find a solution with reasonable values of $\beta$, of the order of  $1/3$ using the BESIII data and of order 0.1 using the LHCb data, with a value around 0.2 for the input of~\cite{Geng:2008gx}. The results obtained clearly indicate that we can understand the ratio $R_B$ in terms of FSI with a strength of the internal emission mechanism in an acceptable range. Certainly we expect that future experimental concerning the $a_0(1710)$ resonance pin down more closely the values of the mass and width of the resonance, which at present have still too much deviation.
\section{Conclusions} 

  The $D_s^+\to \rho^+ \phi$ and $D_s^+\to \rho^+ \omega$ weak decays proceed directly through very different topologies. While the first reaction proceeds via external emission,  the second one does not go in a first step through external nor internal emission, which are the two modes with largest strength. This realization lead the BESIII team to claim that this decay mode proceeds via $W-$annihilation. Then one might expect that the $D_s^+\to \rho^+ \omega$ decay rate is suppressed by at least two orders of magnitude with respect to that of the $D_s^+\to \rho^+ \phi$ one. Yet, this is not the case and the observed suppression is only of a factor about 4. We have then exploited the effect of final state interaction of the pair of vectors originated in the first step of the decay. The external emission directly produces $\rho^+ \phi$, while the internal emission produces directly $K^{*+}\bar{K}^{*0}$.  The studies of the vector vector interaction done in several works show that in isospin $I=1$, one has the interacting channels $\rho \phi, ~\rho \omega$ and $K^{*}\bar{K}^{*}$, and the interaction produces an $a_0$ resonance in the region of 1780 MeV. The resonance was found experimentally later by the Babar, BESIII and LHCb collaborations and the masses are in between 1710 MeV and 1817 MeV, with the theoretical mass in the middle. The original $\rho^+ \phi$ or $K^{*+}\bar{K}^{*0}$ channels make transition to the final $\rho^+ \phi$ and $\rho^+ \omega$, which is led by the dynamically generated $a_0(1710)$ resonance. What we find is that the final state interaction is very important in these processes, and can reduce the original $D_s^+\to \rho^+ \phi$ decays rate appreciably, while it provides a path for the $D_s^+\to \rho^+ \omega$ decay to occur through original production of $K^{*+}\bar{K}^{*0}$ followed by transition to $\rho^+ \omega$. 
  
    We have evaluated the ratio of the two branching ratios and find that, with reasonable values of the relative weight of internal to external emission of the order of $1/N_c$, one can obtain the experimental values of that ratio.  We, thus, demonstrate that the final state interaction of the vector mesons can explain the experimental findings and there is no need to invoke an exceptionally large $W-$annihilation mechanism. At the same time we show in one more example the role of the lately observed $a_0(1710)$ resonance, which is turning into a key ingredient to explain unexpected observables in many new experiments. 
 
\section{Acknowledgments}
This work is partly supported by the National Natural Science
Foundation of China under Grants  No. 12405089 and No. 12247108 and
the China Postdoctoral Science Foundation under Grant
No. 2022M720360 and No. 2022M720359. This work is also supported by
the Spanish Ministerio de Economia y Competitivi-
dad (MINECO) and European FEDER funds under
Contracts No. FIS2017-84038-C2-1-P B, PID2020-
112777GB-I00, and by Generalitat Valenciana under con-
tract PROMETEO/2020/023. This project has received
funding from the European Union Horizon 2020 research
and innovation programme under the program H2020-
INFRAIA-2018-1, grant agreement No. 824093 of the
STRONG-2020 project. This work is supported by the Spanish Ministerio de Ciencia e Innovaci\'on (MICINN) under contracts PID2020-112777GB-I00, PID2023-147458NB-C21 and CEX2023-001292-S; by Generalitat Valenciana under contracts PROMETEO/2020/023 and  CIPROM/2023/59.

\bibliography{refs.bib} 
\end{document}